\UseRawInputEncoding
\documentclass[3p,times,twocolumn]{elsarticle}
 \biboptions{comma,sort&compress}

\usepackage{graphicx}
\usepackage{amsmath}
\usepackage{here}
\usepackage{ecrc}

\volume{00}

\firstpage{1}

\journalname{Nuclear and Particle Physics Proceedings}
\runauth{Bo Wang}

\jid{nppp}
\jnltitlelogo{Nuclear and Particle Physics Proceedings}

\usepackage{amssymb}

\usepackage[figuresright]{rotating}

\begin{document}

\begin{frontmatter}

\title{Molecular tetraquarks and pentaquarks in chiral effective field theory}

\cortext[cor0]{Talk presented at QCD22, 25th International Conference in QCD (4-7, July, 2022,
  Montpellier - FR). }

\author[label1]{Bo Wang}\ead{wangbo@hbu.edu.cn}
\address[label1]{School of Physical Science and Technology, Hebei University, Baoding 071002, China\\
   and\\
Key Laboratory of High-precision Computation and Application of Quantum Field Theory of Hebei Province, Baoding 071002, China}

\author[label2]{Lu Meng}\ead{lu.meng@rub.de}
\address[label2]{Ruhr-Universit\"at Bochum, Fakult\"at f\"ur Physik und Astronomie, Institut f\"ur Theoretische Physik II, D-44780 Bochum, Germany}

\author[label3]{Shi-Lin Zhu}\ead{zhusl@pku.edu.cn}
\address[label3]{School of Physics and Center of High Energy Physics, Peking University, Beijing 100871, China}

\pagestyle{myheadings}
\markright{ }
\begin{abstract}
\noindent
We studied the $D\bar{D}^\ast/\bar{D}D^\ast$, $D^\ast\bar{D}^\ast$, $\bar{D}_s D^\ast/\bar{D}_s^\ast D$, $B\bar{B}^\ast/\bar{B}B^\ast$ and $B^\ast\bar{B}^\ast$ di-hadron interactions in chiral effective field theory ($\chi$EFT) up to the next-to-leading (NLO) order. The above-threshold tetraquark states $Z_c(3900)$, $Z_c(4020)$, $Z_{cs}(3985)$, $Z_b(10610)$, $Z_b(10650)$ can be explained as the corresponding di-hadron resonances, respectively. We also studied the $\Sigma_c^{(\ast)}\bar{D}^{(\ast)}$ interactions to investigate the three hidden-charm pentaquarks $P_c(4312)$, $P_c(4440)$ and $P_c(4457)$. We further used the parameters fixed from the $P_c$ states to predict the possible molecular states in $\Xi_c\bar{D}^{(\ast)}$, $\Xi_c^\prime\bar{D}^{(\ast)}$ and $\Xi_c^\ast\bar{D}^{(\ast)}$ systems. Our predictions of the $\Xi_c\bar{D}^{(\ast)}$ bound states are very consistent with two new near-threshold structures recently observed by the LHCb Collaboration.

\end{abstract}
\end{frontmatter}
\section{Introduction}
In past decades, many near-threshold exotic hadrons were observed in experiments~\cite{Workman:2022ynf,Chen:2016qju,Guo:2017jvc,Liu:2019zoy,Brambilla:2019esw,Meng:2022ozq}. The near-threshold behaviors of these states indicate that they are good candidates of the hadronic molecules that are composed of the corresponding di-hadrons. These possible molecular states provide good platforms to study the hadronic molecular dynamics. A rational conjecture is that the interaction between the heavy-light systems is just a duplication of the few-body nuclear forces. Based on this conjecture, the heavy flavor molecular states were the cousins of the deuteron. I.e., they are the deuteron-like heavy hadronic molecules. In our previous studies~\cite{Meng:2019ilv,Wang:2019ato,Wang:2019nvm,Wang:2020dko,Wang:2020htx,Meng:2019nzy,Wang:2020dhf}, we used the $\chi$EFT to investigate the interactions of the paired heavy-light systems. In these systems, both the chiral symmetry and heavy quark symmetry are manifested~\cite{Meng:2022ozq}.

\begin{figure}[!hptb]
\begin{centering}
    \scalebox{1.0}{\includegraphics[width=0.8\linewidth]{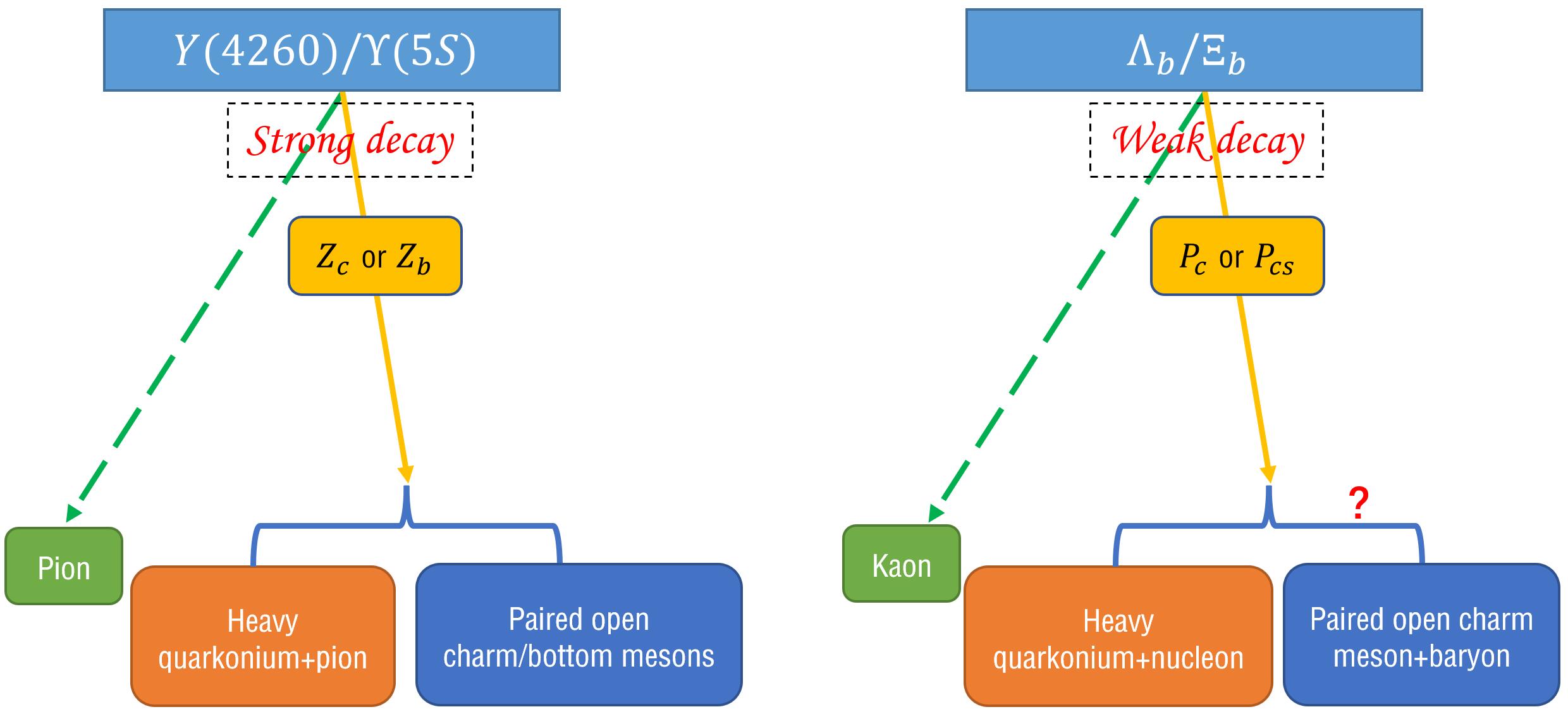}}
    \caption{An illustration of the production processes for $Z_{c,b}$ and $P_{c(s)}$ states in experiments. The open-charm decay channels of $P_{c(s)}$ states are unobserved yet, so the corresponding decay route is marked by the question mark.\label{fig:fig1}}
\end{centering}
\end{figure}

We first briefly overview the experimental observations of the states that we focused on in this report. The $Z_c(3900)$ and $Z_c(4020)$ were observed by the BESIII Collaboration in the $J/\psi\pi$~\cite{BESIII:2013ris} and $h_c\pi$~\cite{BESIII:2013ouc} channels, respectively. They were also confirmed in the open-charm channels, i.e., the $D\bar{D}^\ast/\bar{D}D^\ast$~\cite{BESIII:2013qmu} and $D^\ast\bar{D}^\ast$~\cite{BESIII:2013mhi}, respectively. Recently, the BESIII reported a structure, the $Z_{cs}(3985)$ in $e^+e^-\to K^+(D_s^- D^{\ast0}+D_s^{\ast-} D^0)$ process~\cite{BESIII:2020qkh} (the evidence of the neutral one was also reported in Ref.~\cite{BESIII:2022qzr}). The $Z_b(10610)$ and $Z_b(10650)$ were observed by the Belle Collaboration in the $\Upsilon(nS)\pi~(n=1,2,3)$ and $h_b(mP)\pi~(m=1,2)$ invariant mass spectra~\cite{Belle:2011aa}, and they were also confirmed in the open-bottom channels $B\bar{B}^\ast/\bar{B}B^\ast$ and $B^\ast\bar{B}^\ast$~\cite{Belle:2015upu}, respectively. Three pentaquarks $P_c(4312)$, $P_c(4440)$ and $P_c(4457)$ were observed by the LHCb Collaboration in the $J/\psi p$ channel~\cite{LHCb:2019kea}. The mass, width, the measured$/$preferred $I(J^P)$ quantum numbers, the nearest threshold and the observed channels of these states are listed in Table~\ref{tab:tab1}, and the corresponding production processes are illustrated in Fig~\ref{fig:fig1}.
\begin{table*}[htbp]
\centering
\renewcommand{\arraystretch}{1.1}
\caption{The mass, width, the measured$/$preffered $I(J^P)$ quantum numbers, the nearest threshold and the observed channels of the considered states.\label{tab:tab1}}
\setlength{\tabcolsep}{1.9mm}
{
\begin{tabular}{cccccc}
\hline
State& Mass (MeV)& Width (MeV)& $I(J^P)$& Nearest threshold& Observed channels\\
$Z_c(3900)$& $3887.1\pm2.5$ & $28.4\pm2.6$ & $1(1^+)$ & $D\bar{D}^\ast/\bar{D}D^\ast$ & $J/\psi\pi$, $D\bar{D}^\ast/\bar{D}D^\ast$~\cite{BESIII:2013ris,BESIII:2013qmu}\\
$Z_c(4020)$& $4024.1\pm1.9$ & $13\pm5$ & $1(1^+)$ & $D^\ast\bar{D}^\ast$ & $h_c\pi$, $D^\ast\bar{D}^\ast$~\cite{BESIII:2013ouc,BESIII:2013mhi}\\
$Z_{cs}(3985)$& $3982.5^{+1.8}_{-2.6}\pm2.1$ & $12.8^{+5.3}_{-4.4}\pm3.0$ & $\frac{1}{2}(1^+)$ & $\bar{D}_s D^\ast/\bar{D}_s^\ast D$ & $\bar{D}_s D^\ast/\bar{D}_s^\ast D$~\cite{BESIII:2020qkh}\\
$Z_b(10610)$& $10607.2\pm2.0$ & $18.4\pm2.4$ & $1(1^+)$ & $B\bar{B}^\ast/\bar{B}B^\ast$ & $\Upsilon(nS)\pi$, $h_b(mP)\pi$, $B\bar{B}^\ast/\bar{B}B^\ast$~\cite{Belle:2011aa,Belle:2015upu}\\
$Z_b(10650)$& $10652.2\pm1.5$ & $11.5\pm2.2$ & $1(1^+)$ & $B^\ast\bar{B}^\ast$ & $\Upsilon(nS)\pi$, $h_b(mP)\pi$, $B^\ast\bar{B}^\ast$~\cite{Belle:2011aa,Belle:2015upu}\\
$P_c(4312)$& $4311.9\pm0.7^{+6.8}_{−0.6}$& $9.8\pm2.7^{+3.7}_{−4.5}$& $\frac{1}{2}(\frac{1}{2}^-)$& $\Sigma_c\bar{D}$& $J/\psi p$~\cite{LHCb:2019kea}\\
$P_c(4440)$& $4440.3\pm1.3^{+4.1}_{−4.7}$& $20.6\pm4.9^{+8.7}_{−10.1}$& $\frac{1}{2}(\frac{1}{2}^-)$& $\Sigma_c\bar{D}^\ast$& $J/\psi p$~\cite{LHCb:2019kea}\\
$P_c(4457)$& $4457.3\pm0.6^{+4.1}_{−1.7}$& $6.4\pm2.0^{+5.7}_{−1.9}$& $\frac{1}{2}(\frac{3}{2}^-)$& $\Sigma_c\bar{D}^\ast$& $J/\psi p$~\cite{LHCb:2019kea}\\
\hline
\end{tabular}
}
\end{table*}

\section{A short introduction to the $\chi$EFT}

The $\chi$EFT inherits the general features of EFT who works at low energies. For example, the high energy dynamics are integrated out such that their contributions can be mimicked by the four-body contact interactions. The contributions from the soft degrees of freedom can be calculated to any given order with the standard procedures in perturbative quantum field theory. The applications of $\chi$EFT have achieved lots of progresses in few-body nuclear systems in the past decades~\cite{Weinberg:1990rz,Weinberg:1991um,Epelbaum:2008ga,Machleidt:2011zz}. The generalization of the $\chi$EFT to the two-body heavy-light systems is straightforward once their interactions are constrained by the chiral and heavy quark symmetries~\cite{Meng:2022ozq}.

In $\chi$EFT, the nonrelativistic scattering amplitudes of two heavy matter fields can be expanded in powers of a small parameter $\mathcal{Q}/\Lambda$, where $\mathcal{Q}$ generally denotes the soft scale, such as the pion mass or the transferred momenta, while the $\Lambda$ represents the hard scale, such as the vector ($\rho$) meson's mass or the chiral symmetry breaking scale $\Lambda_\chi\sim1$ GeV. If one considers the nonrelativistic correction terms, then $\mathcal{Q}$ could be the mass splittings of the spin multiplets (such as the mass splittings of $D$, $D^\ast$), and $\Lambda$ could be the masses of the heavy matter fields.

\begin{figure}[!hptb]
\begin{centering}
    \scalebox{1.0}{\includegraphics[width=0.8\linewidth]{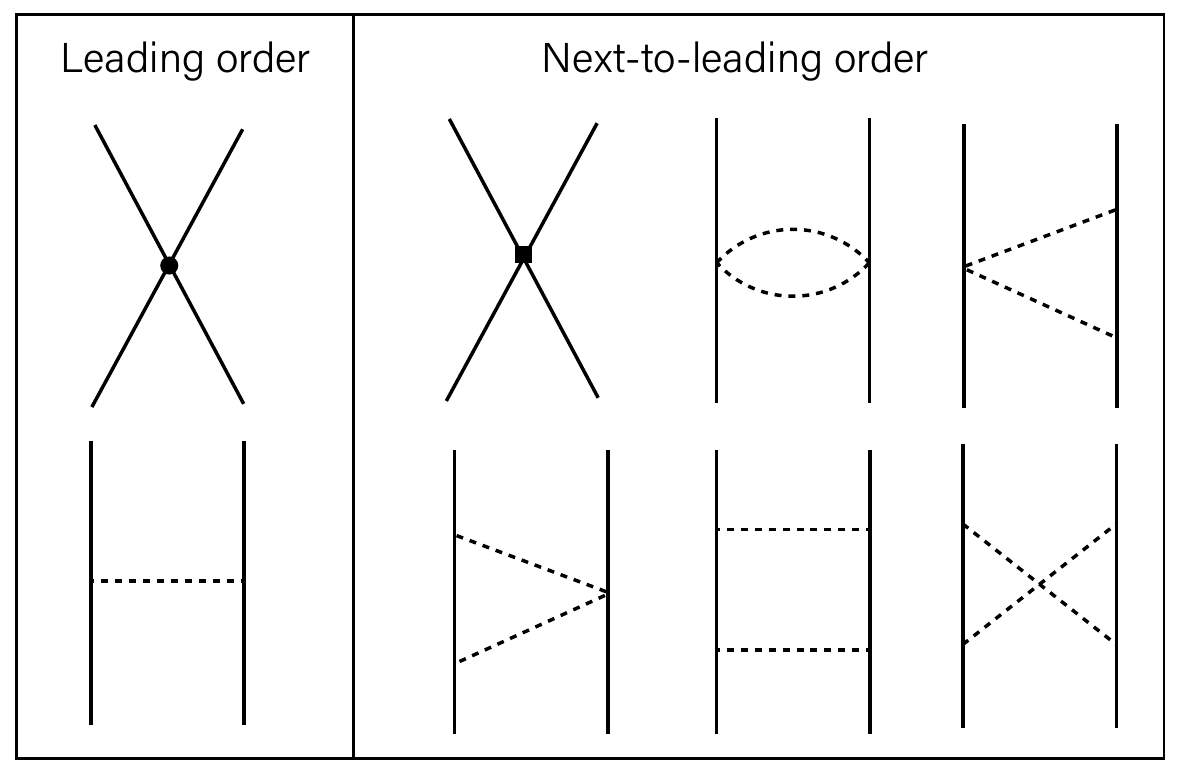}}
    \caption{The topological Feynman diagrams that contribute to the scatterings of two heavy-light hadrons in the framework of $\chi$EFT up to the NLO. The solid and dashed lines denote the matter fields and pion, respectively.\label{fig:fig2}}
\end{centering}
\end{figure}

Generally, the scattering amplitudes are written as
\begin{eqnarray}\label{eq:amp}
\mathcal{M}&=&\sum_\nu\mathcal{C}\left(\frac{\mathcal{Q}}{\mu},g_i\right)\left(\frac{\mathcal{Q}}{\Lambda}\right)^\nu,
\end{eqnarray}
where $\mu$ is the regularization scale that appears in the loop diagram calculations, and $g_i$ stands for the coupling constants in the Lagrangians. The power $\nu$ is related to the power counting. One only needs to calculate finite (sometimes few) terms for a given accuracy if the convergence of the expansion behaves well. The terms with $\nu=0$ and $2$ correspond to the leading order (LO) and NLO ones, respectively. The corresponding Feynman diagrams are shown in Fig.~\ref{fig:fig2}.

With the amplitudes in Eq.~\eqref{eq:amp}, one can derive the effective potentials. These effective potentials can be iterated into the Schr\"odinger equation or the Lippman-Schwinger equation (LSE) to search for the possible bound states or resonances. Usually, it is more convenient to use the momentum-space LSE to study the experimental line-shapes. One should note that the effective potentials were always regularized via introducing a finite cutoff.

\begin{figure*}[htbp]
\begin{center}
\begin{minipage}[t]{0.23\linewidth}
\centering
\includegraphics[width=\columnwidth]{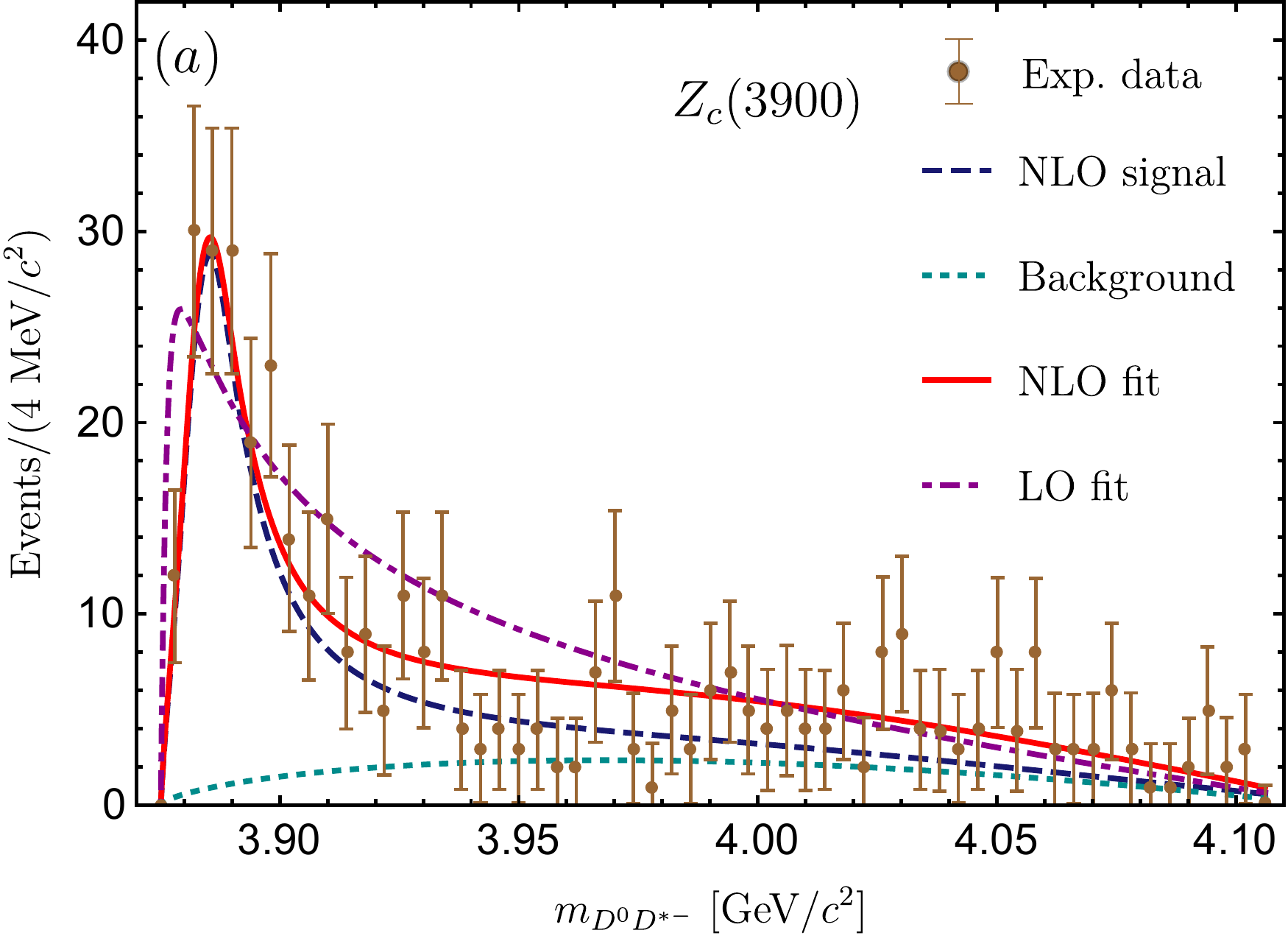}
\end{minipage}%
\begin{minipage}[t]{0.23\linewidth}
\centering
\includegraphics[width=\columnwidth]{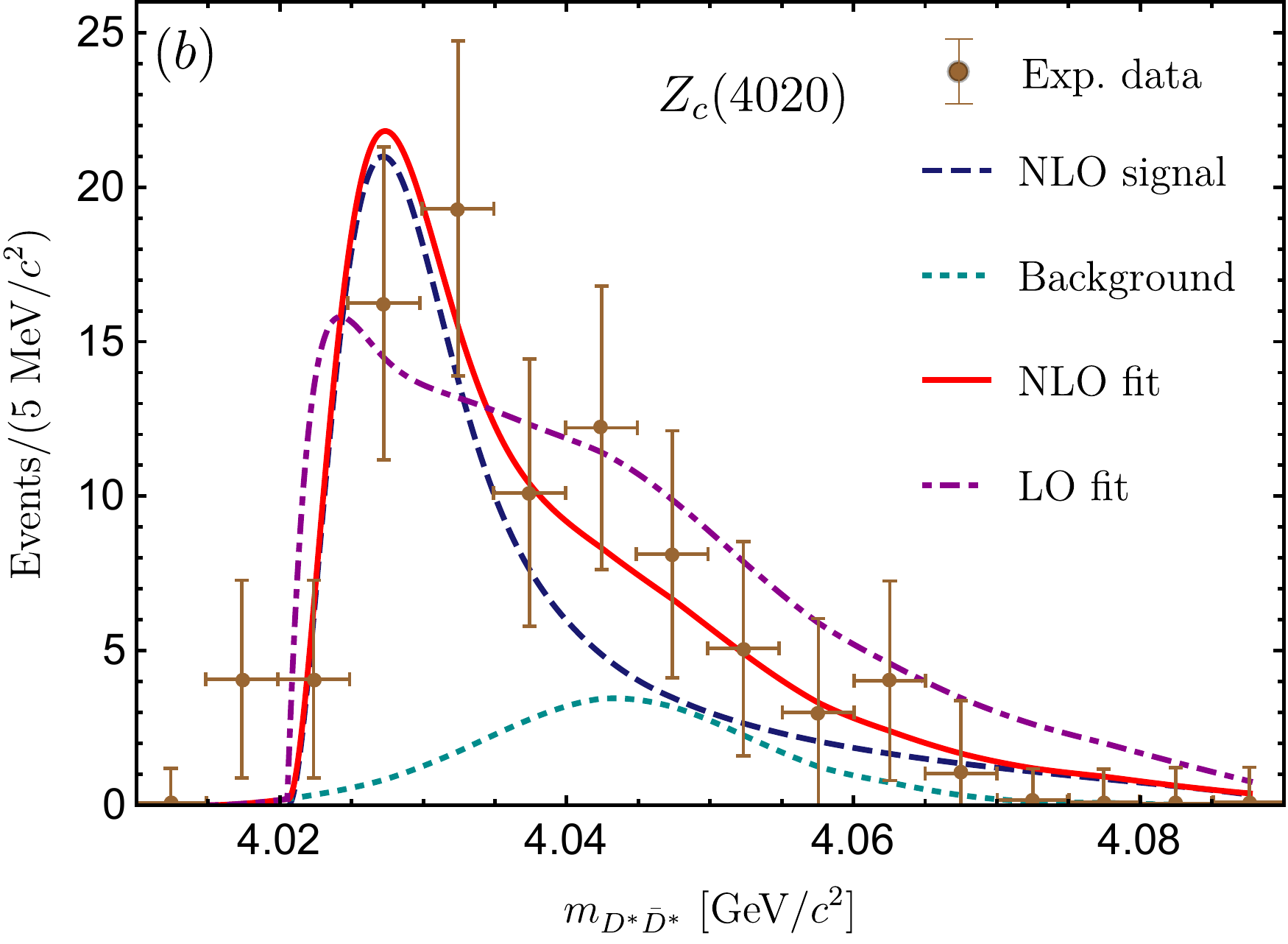}
\end{minipage}
\begin{minipage}[t]{0.242\linewidth}
\centering
\includegraphics[width=\columnwidth]{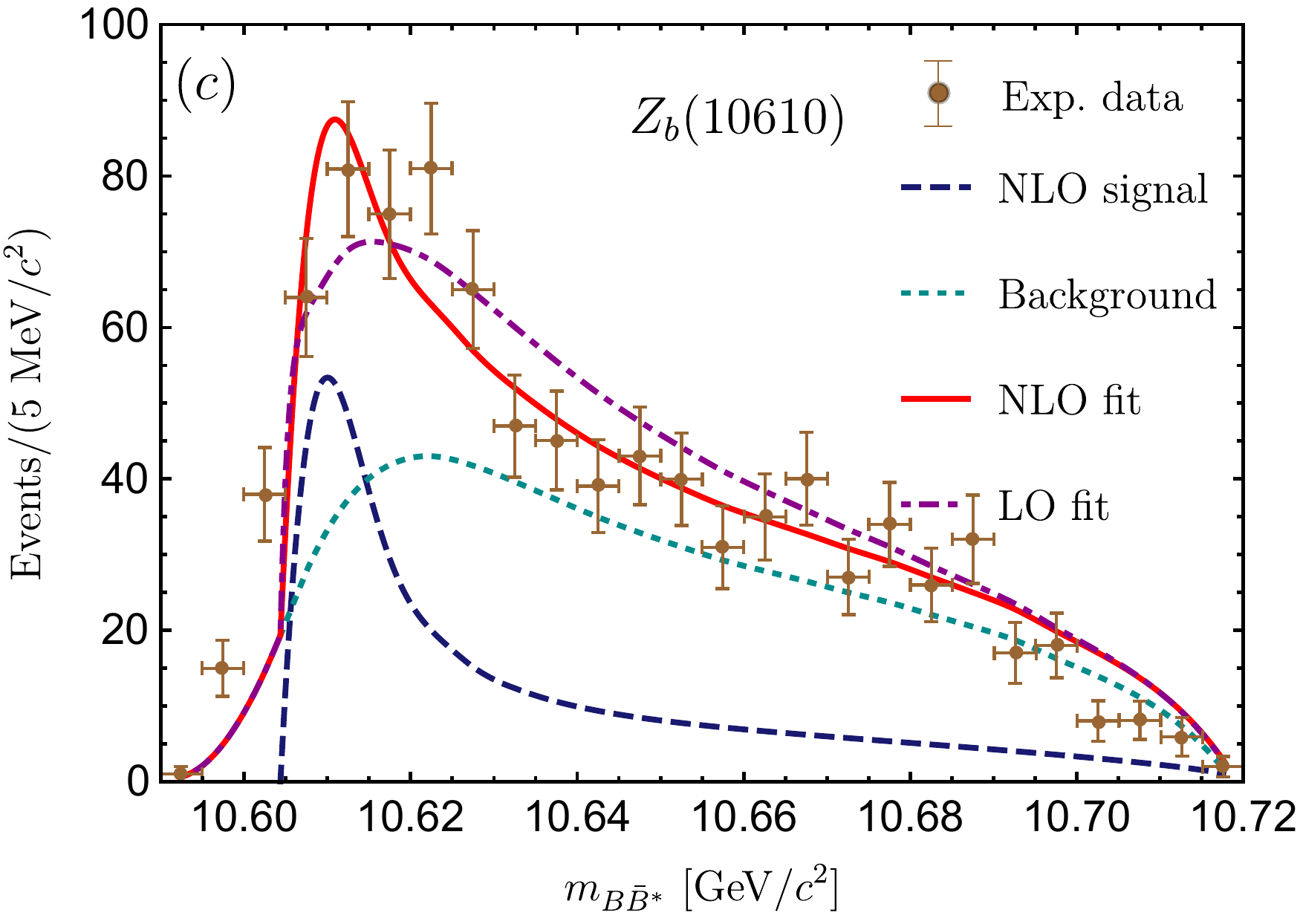}
\end{minipage}%
\begin{minipage}[t]{0.24\linewidth}
\centering
\includegraphics[width=\columnwidth]{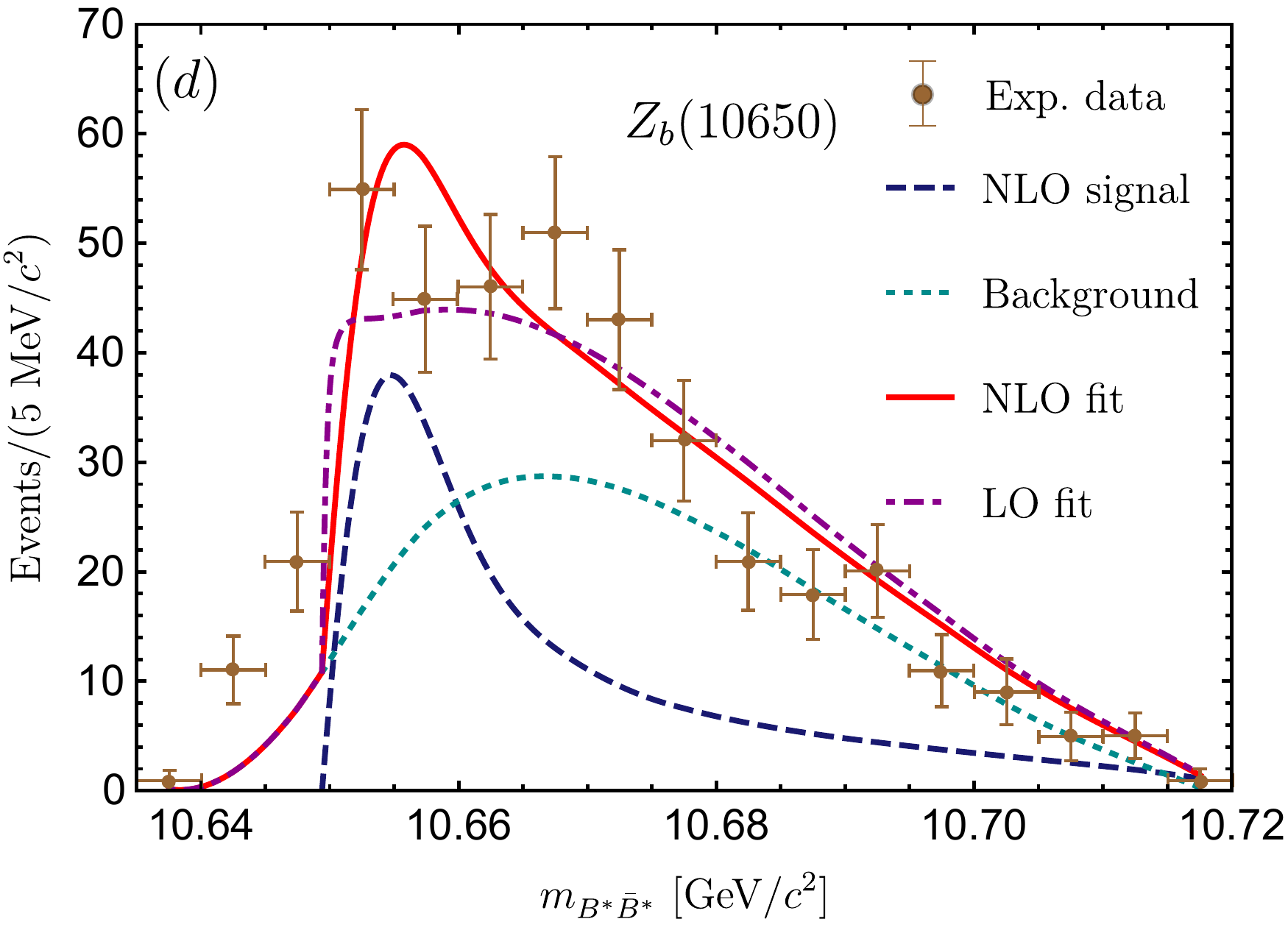}
\end{minipage}
\caption{The fitted invariant mass spectra of the $D\bar{D}^\ast/\bar{D}D^\ast$, $D^\ast\bar{D}^\ast$, $B\bar{B}^\ast/\bar{B}B^\ast$ and $B^\ast\bar{B}^\ast$ for the $Z_c(3900)$, $Z_c(4020)$, $Z_b(10610)$ and $Z_b(10650)$, respectively. The notations are marked in each subfigure.\label{fig:fig3}}
\end{center}
\end{figure*}

\section{Molecular tetraquarks and pentaquarks}

\subsection{$Z_{c(s)}$ and $Z_{b(s)}$ states}

In Ref.~\cite{Wang:2020dko}, we simultaneously studied the experimental line-shapes of $Z_c(3900)$, $Z_c(4020)$, $Z_b(10610)$ and $Z_b(10650)$. We considered the production processes $\gamma^\ast\to\mathtt{V}\mathtt{P}(\mathtt{V})\pi$ [where $\gamma^\ast$ denotes a virtual photon with center of mass energy $\sqrt{S}$, $\mathtt{V}(\mathtt{P})$ denotes the vector (pseudoscalar) charmed/bottom mesons], and the rescattering T-matrix of $\mathtt{V}\mathtt{P}(\mathtt{V})$ systems are attached to the production vertex. The undetermined low energy constants (LECs) were fitted from the $\mathtt{V}\mathtt{P}(\mathtt{V})$ invariant mass spectra. The results are shown in Fig.~\ref{fig:fig3}. One can see that the experimental data can be well fitted when using the LO plus NLO potentials.

In our study, we did not consider the hidden-charm (bottom) channel contributions. On the one hand, we found their contributions to the formations of the $Z_{c,b}$ states are marginal, i.e., they may mainly serve as the decay products rather than the inner components of the $Z_{c,b}$ states. On the other hand, these hidden-charm (bottom) channels lie far below the thresholds of $Z_{c,b}$ states, and it is hard to seriously take them into account in the framework of $\chi$EFT. 

We further studied the recently observed $Z_{cs}(3985)$ via treating it as the SU(3) partner of $Z_c(3900)$~\cite{Wang:2020htx,Meng:2020ihj}. The values for the LECs of $\bar{D}_s D^\ast/\bar{D}_s^\ast D$ systems are from the fitted ones of $Z_c(3900)$. We noticed that the $\bar{D}_s D^\ast/\bar{D}_s^\ast D$ invariant mass spectrum and the corresponding mass and width of $Z_{cs}(3985)$ can all be well reproduced~\cite{Wang:2020htx}.

\begin{figure}[!htbp]
\begin{center}
\begin{minipage}[t]{0.45\linewidth}
\centering
\includegraphics[width=\columnwidth]{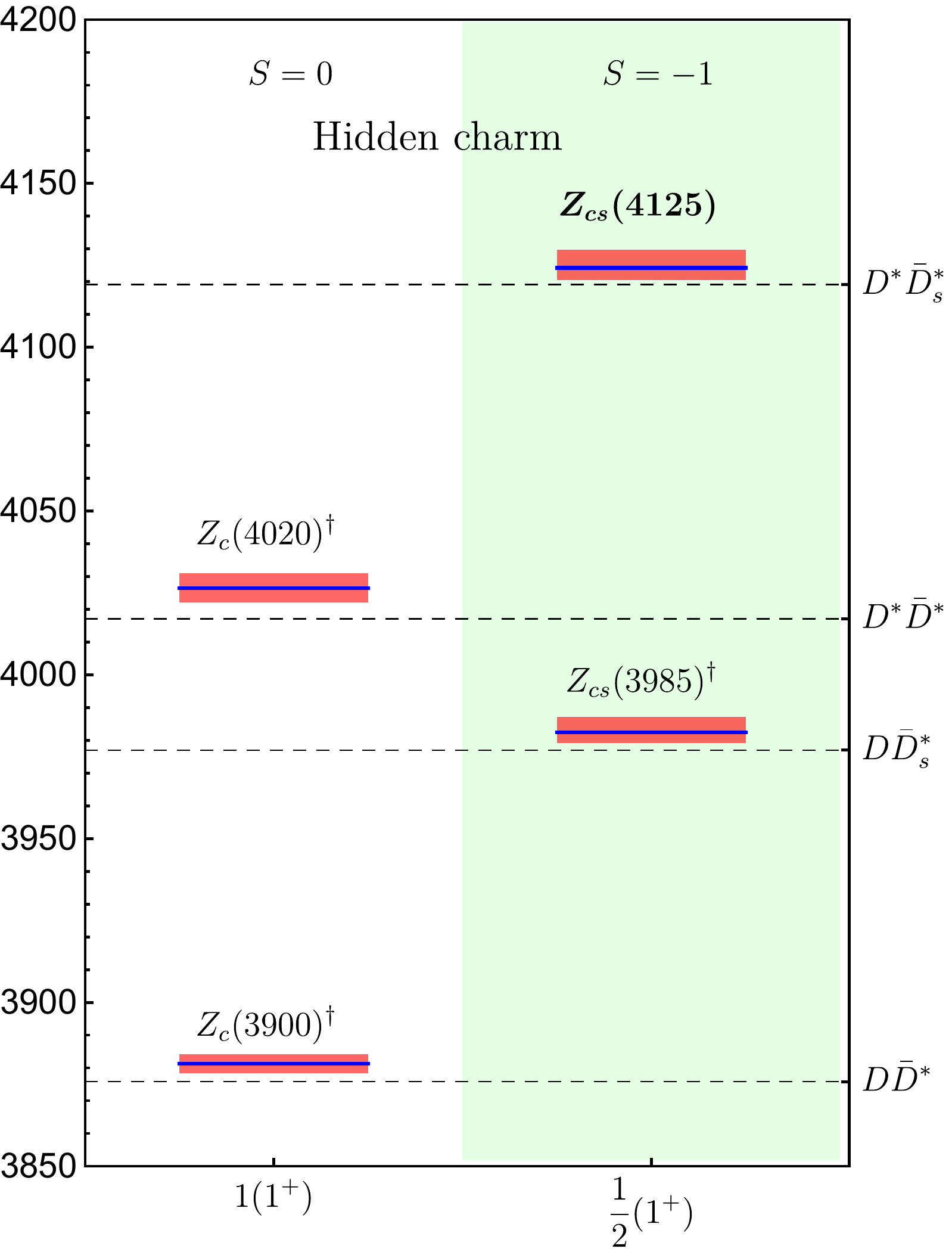}
\end{minipage}%
\begin{minipage}[t]{0.46\linewidth}
\centering
\includegraphics[width=\columnwidth]{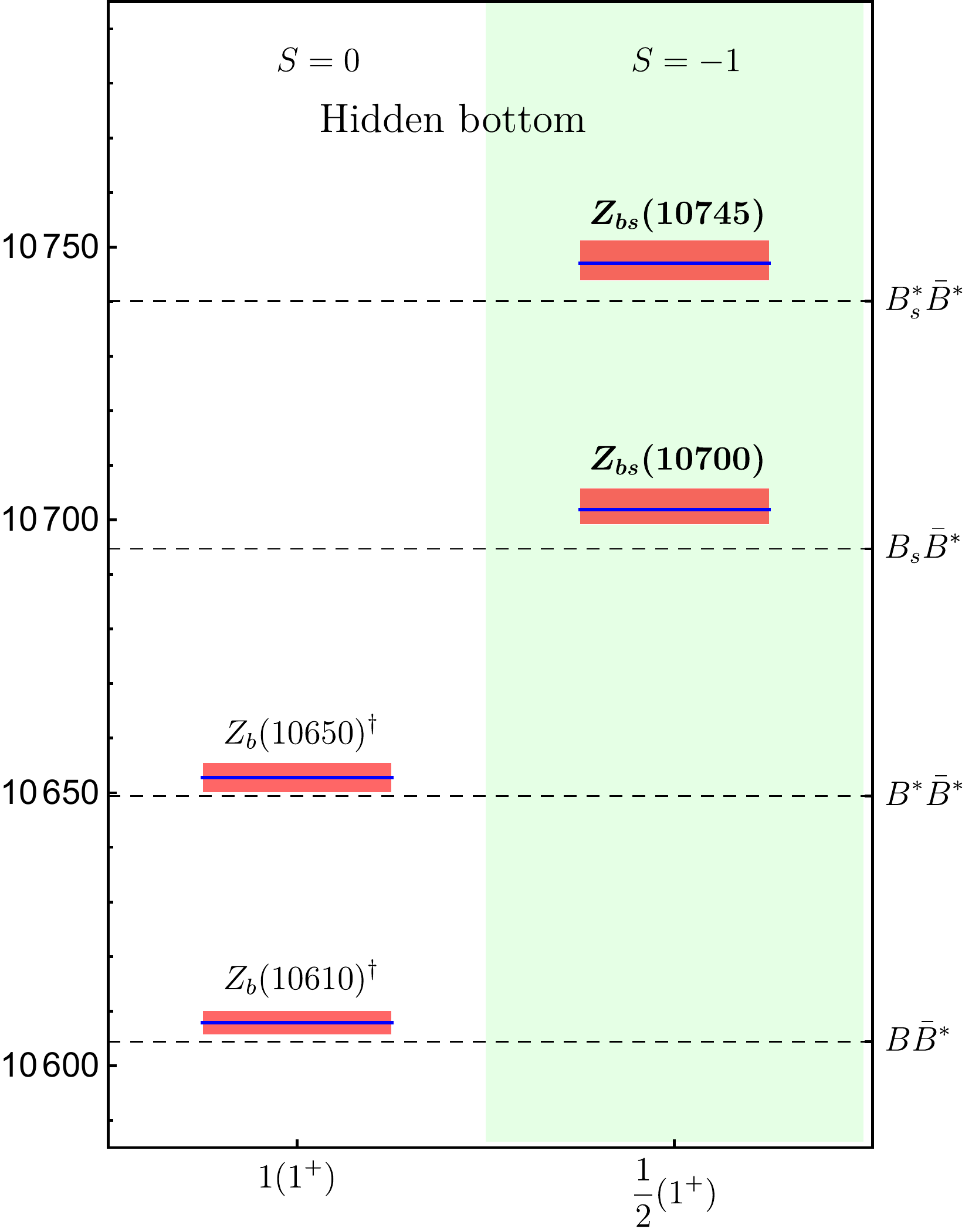}
\end{minipage}
\caption{The mass spectrum of the hidden-charm and hidden-bottom molecular tetraquarks with strangeness $S=0$ and $S=-1$, respectively. The observed and predicted states are marked with `$\dagger$' and boldface, respectively.\label{fig:fig4}}
\end{center}
\end{figure}

We established a complete mass spectrum of the charged charmoniumlike and bottomoniumlike states with strangeness $S=0$ and $S=-1$, see Fig.~\ref{fig:fig4}. We found the masses of these states all lie above the corresponding di-hadron thresholds, and the poles of the T-matrix reside in the second Riemann sheet, which implies that they are the resonances rather than the bound states or virtual states.

\begin{figure*}[!htbp]
\begin{center}
\begin{minipage}[t]{0.242\linewidth}
\centering
\includegraphics[width=\columnwidth]{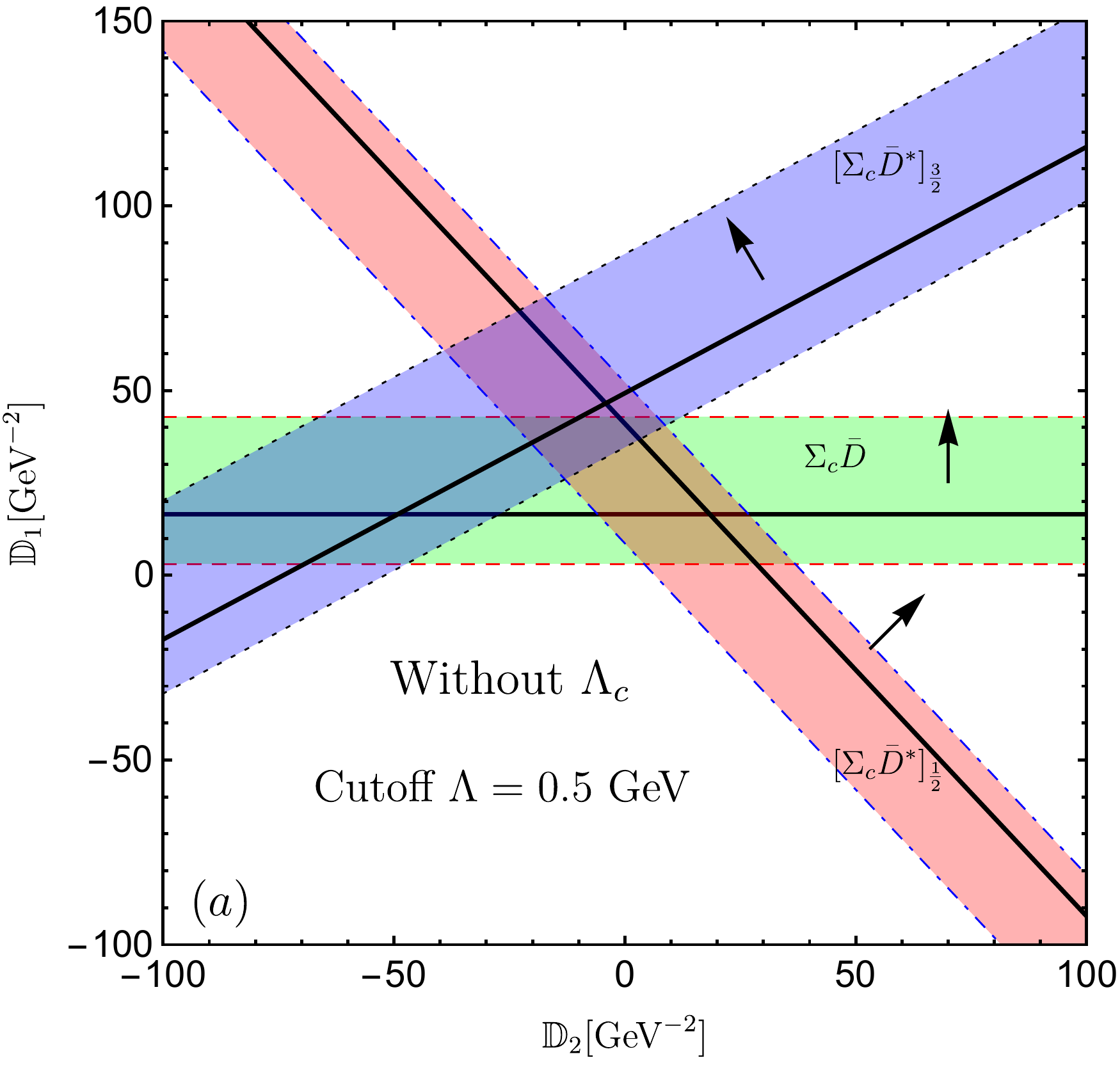}
\end{minipage}%
\begin{minipage}[t]{0.242\linewidth}
\centering
\includegraphics[width=\columnwidth]{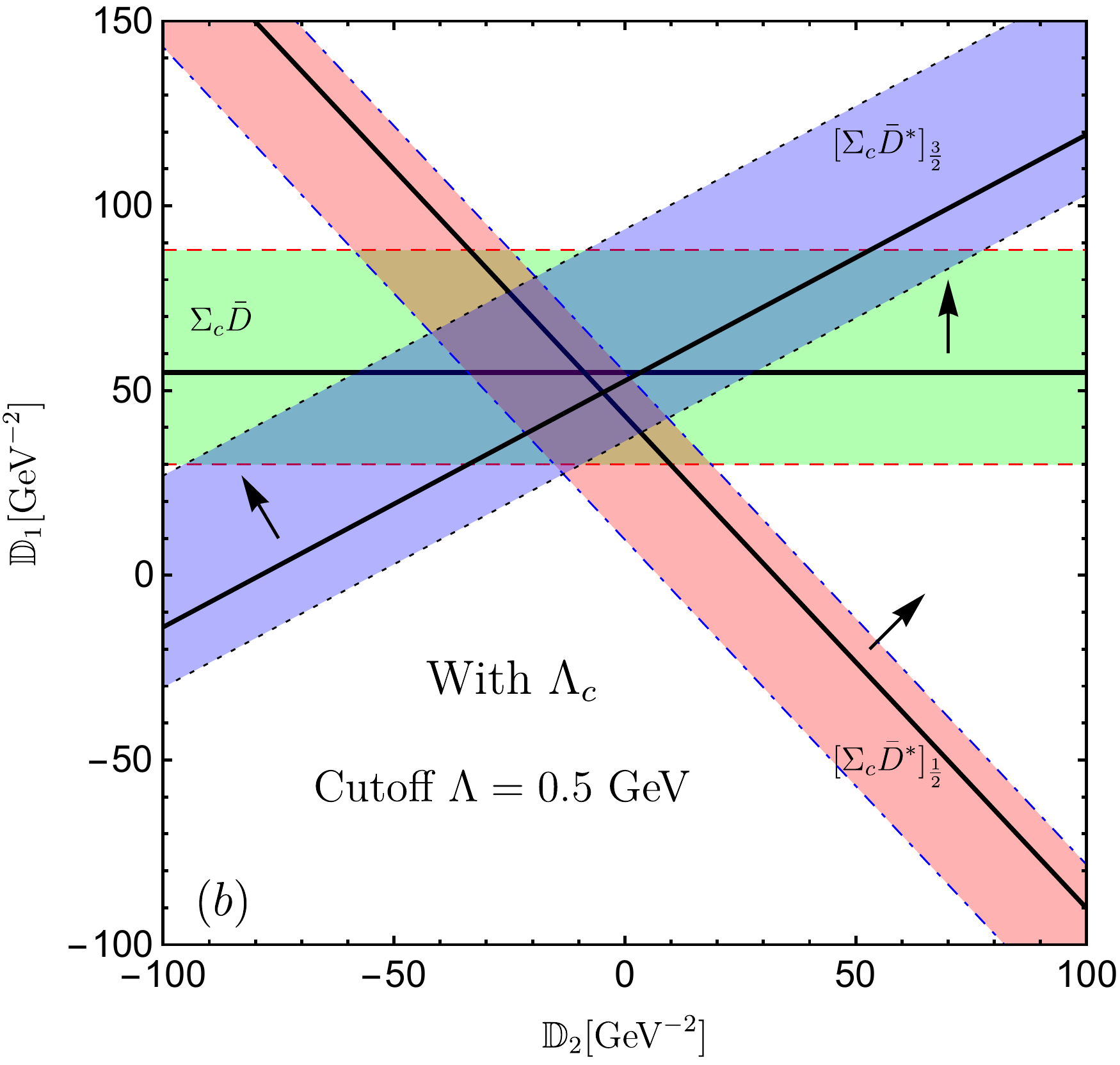}
\end{minipage}
\begin{minipage}[t]{0.242\linewidth}
\centering
\includegraphics[width=\columnwidth]{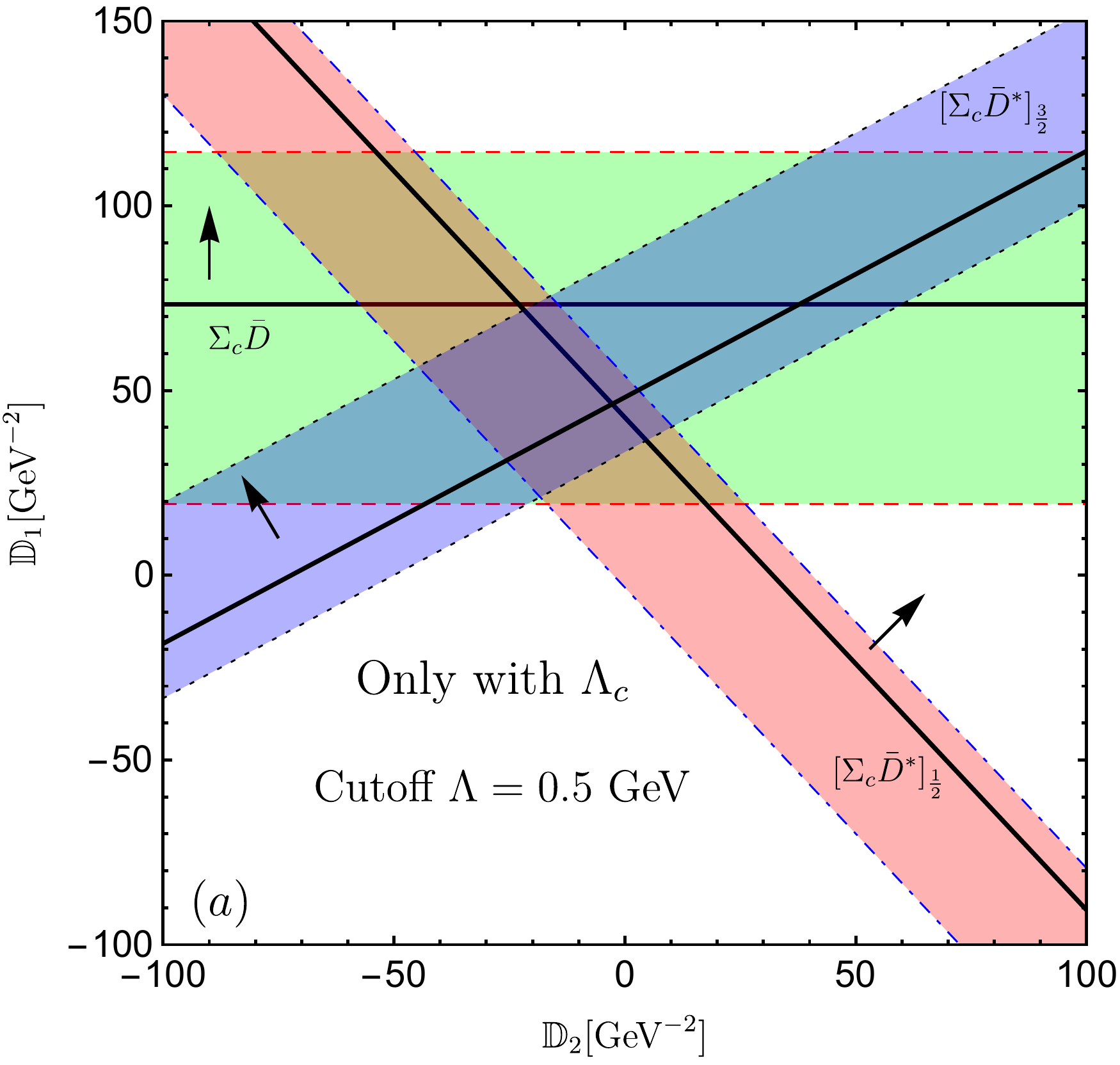}
\end{minipage}%
\begin{minipage}[t]{0.242\linewidth}
\centering
\includegraphics[width=\columnwidth]{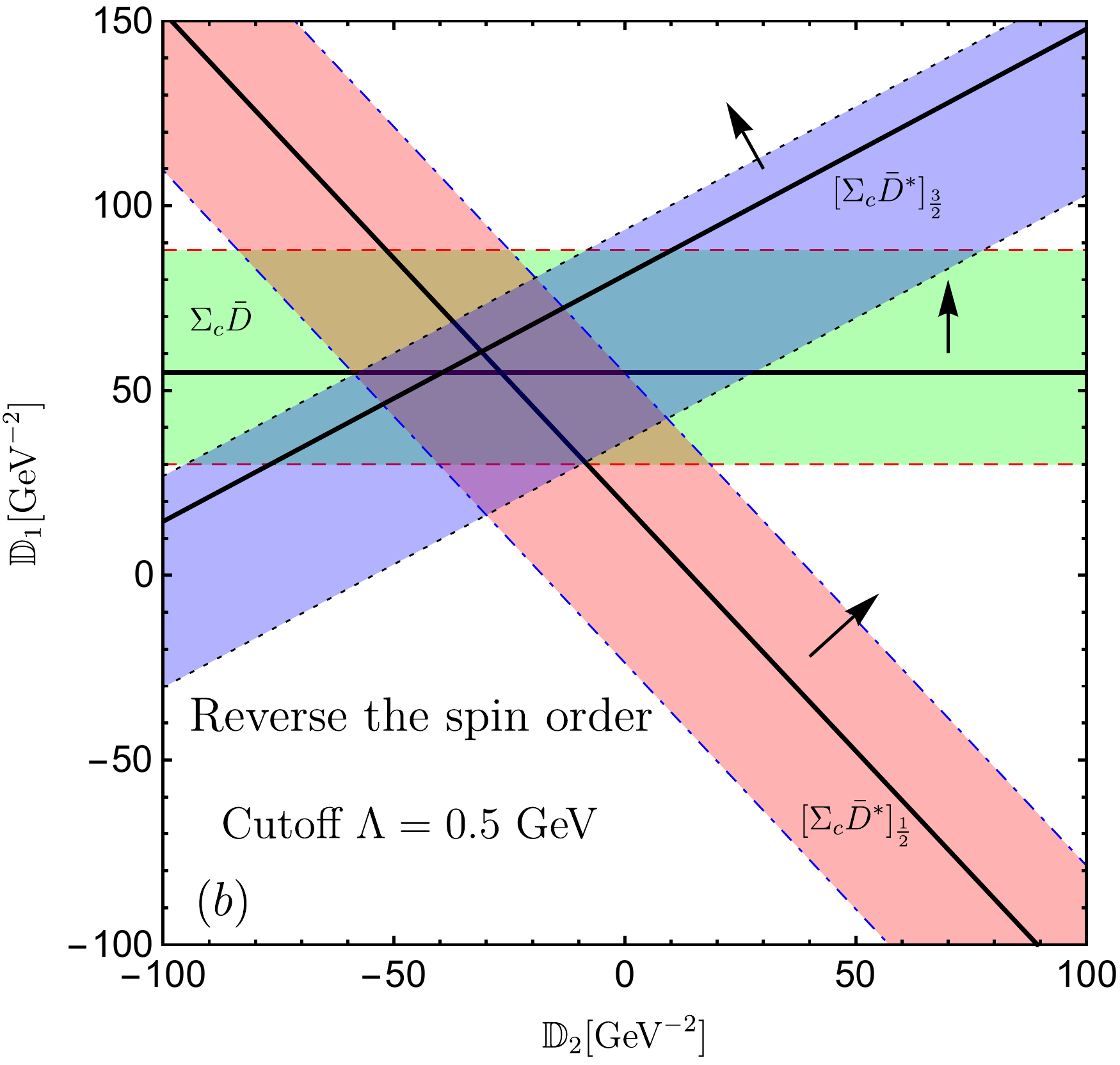}
\end{minipage}
\caption{The fitted results for $P_c(4312)$, $P_c(4440)$ and $P_c(4457)$ in four different cases. The result would be better if the three black solid lines tend to intersect at one point. More detailed explanations can be found in Ref.~\cite{Wang:2019ato}.\label{fig:fig5}}
\end{center}
\end{figure*}

\subsection{$P_c$ and $P_{cs}$ states}

We also studied the $\Sigma_c\bar{D}^{(\ast)}$ interactions within the $\chi$EFT~\cite{Wang:2019ato}, because the bound states in these systems may correspond to the $P_c(4312)$, $P_c(4440)$ and $P_c(4457)$.

There are two free parameters---the two LO LECs $\mathbb{D}_1$ and $\mathbb{D}_2$, which denote the strength of the central term and spin-spin interaction, respectively. In our study, we fixed the cutoff at $0.5$ GeV, and varied the $\mathbb{D}_1$ and $\mathbb{D}_2$ in the range $[-100,150]$ GeV$^{-2}$ and $[-100,100]$ GeV$^{-2}$, respectively. We considered four different cases: (1) We did not consider the contribution of the $\Lambda_c$ in the two-pion exchange (TPE) loop diagrams, and the result in this case is not good (the first figure of Fig.~\ref{fig:fig5}). (2) We added the $\Lambda_c$ in the TPE, and we can simultaneously reproduce the three $P_c$ states (the second figure of Fig.~\ref{fig:fig5}). (3) We also tested the result that only considers the $\Lambda_c$ as the intermediate state in TPE, but the result in this case is also unsatisfactory (the third figure of Fig.~\ref{fig:fig5}). (4) In above three cases, we assumed the spins of the $P_c(4312)$, $P_c(4440)$ and $P_c(4457)$ are $\frac{1}{2}$, $\frac{1}{2}$, $\frac{3}{2}$, respectively. If $P_c(4440)$ and $P_c(4457)$ are the S-wave molecules of $\Sigma_c\bar{D}^\ast$, their spin configurations are undetermined. So in this case, we assume the spins of $P_c(4440)$ and $P_c(4457)$ are $\frac{3}{2}$ and $\frac{1}{2}$, respectively. The result is as good as that of case 2 at the price of largely enhancing the spin-spin interactions (the fourth figure of Fig.~\ref{fig:fig5}).



With the input of $P_c$ states, we bridged the LECs of $\Sigma_c^{(\ast)}\bar{D}^{(\ast)}$ to those of $\Xi_c\bar{D}^{(\ast)}/\Xi_c^\prime\bar{D}^{(\ast)}/\Xi_c^\ast\bar{D}^{(\ast)}$ systems with a contact Lagrangian at the quark level~\cite{Wang:2019nvm}. We predicted ten bound states in the isoscalar channels of these systems, see Table III in Ref.~\cite{Wang:2019nvm}. The predicted masses of the two states in $\Xi_c\bar{D}^\ast$ and $\Xi_c\bar{D}$ systems with $J=\frac{1}{2}$ are $4456.9^{+3.2}_{-3.3}$ and $4319.4^{+2.8}_{-3.0}$ MeV, respectively. The results are consistent with two recently observed structures $P_{cs}(4459)$ and $P_{cs}(4338)$ from the LHCb in the $J/\psi\Lambda$ channel near the $\Xi_c\bar{D}^\ast$~\cite{LHCb:2020jpq} and $\Xi_c\bar{D}$~\cite{LHCb:2022pzn} thresholds, respectively.

\section{A short summary}

The $\chi$EFT is a powerful tool in dealing with the near-threshold hadronic molecules.

The $Z_c(3900)$, $Z_c(4020)$, $Z_{cs}(3985)$, $Z_b(10610)$ and $Z_b(10650)$ can be well interpreted as the $D\bar{D}^\ast/\bar{D}D^\ast$, $D^\ast\bar{D}^\ast$, $\bar{D}_s D^\ast/\bar{D}_s^\ast D$, $B\bar{B}^\ast/\bar{B}B^\ast$ and $B^\ast\bar{B}^\ast$ di-hadron resonances, respectively. The $P_c(4312)$, $P_c(4440)$ and $P_c(4457)$ can be well described as the $\Sigma_c\bar{D}$, $\Sigma_c\bar{D}^\ast$ bound states with spin $J=\frac{1}{2},\frac{1}{2}$ and $\frac{3}{2}$, respectively. The $P_{cs}(4338)$ and $P_{cs}(4459)$ may correspond to the predicted states in the $\Xi_c\bar{D}$ and $\Xi_c\bar{D}^\ast$ channels, respectively.



\end{document}